\documentclass[onecolumn]{pasj00} %proof, onecolumn, twocolumn
\SetRunningHead{Hideki Yahagi Yahagi}{Vectorization and
  Parallelization of AMR $N$-body code}

\title{Vectorization and Parallelization of the Adaptive Mesh Refinement $N$-body Code}
\author{Hideki \textsc{Yahagi}
}
\affil{%
  Department of Astronomy, University of Tokyo 
  7--3--1 Hongo, Bunkyo ward, Tokyo 113-0033
}
\email{hyahagi@astron.s.u-tokyo.ac.jp}
\Received{2005 April 13}
\Accepted{$\langle$acception date$\rangle$}
\Published{$\langle$publication date$\rangle$}
\begin{document}
\maketitle

\begin{abstract}
In this paper, we describe our vectorized and parallelized
adaptive mesh refinement (AMR) $N$-body code with shared time steps,
and report its performance on a Fujitsu VPP5000 vector-parallel
supercomputer.  
Our AMR $N$-body code puts hierarchical meshes recursively where
higher resolution is required and the time step of all particles are
the same.
The parts which are the most difficult to vectorize
are loops that access the mesh data and particle data.  We vectorized
such parts by changing the loop structure, so that the innermost loop
steps through the cells instead of the particles in each cell, in
other words, by changing the loop order from the depth-first order to
the breadth-first order.  Mass assignment is also vectorizable using
this loop order exchange and splitting the loop into $2^{N_{\mathrm{dim}}}$
loops, if the cloud-in-cell scheme is adopted. Here, $N_{\mathrm{dim}}$ is the
number of dimension.  These vectorization schemes which eliminate the
unvectorized loops are applicable to parallelization of loops for
shared-memory multiprocessors.  We also parallelized our code for
distributed memory machines.  The important part of parallelization is
data decomposition.  We sorted the hierarchical mesh data by the
Morton order, or the recursive N-shaped order, level by level and
split and allocated the mesh data to the processors.  Particles are
allocated to the processor to which the finest refined cells including
the particles are also assigned.
Our timing analysis using the $\Lambda$-dominated cold dark matter
simulations shows that our parallel code speeds up almost ideally up to
32 processors, the largest number of processors in our test.
\end{abstract}

% main text
%##################################################
% Introduction
%##################################################
\section{Introduction}

By introduction and invention of the high resolution cosmological
$N$-body methods,
such as the particle-particle--particle-mesh (P$^3$M) method
\citep{efse81, hoce88}, cosmological $N$-body simulations became
fundamental and powerful tools to investigate the structure formation
processes in the universe (e.g. \authorcite{davefw85}
\yearcite{davefw85}).  Besides,
parallelization schemes for the P$^3$M method were developed by some
authors \citep{ferb94, ferb95, maccpp98}.  Since the P$^3$M method is
memory efficient, the number of particles in P$^3$M simulations
reached to 10$^9$ \citep{evretal02}.  However, the P$^3$M method is
not almighty.  In the P$^3$M method, forces from nearby particles are
calculated by the direct sum method.  Hence, as the system gets
clustered, direct sum part dominates the whole CPU time and the
computational cost grows rapidly.  In order to resolve this defect of
the P$^3$M method, some methods are proposed, such as the adaptive
P$^3$M method \citep{cou91} which uses adaptive meshes in the dense
regions, and the tree particle-mesh method \citep{xu95, bag02, bodo03}
which uses tree method \citep{barh85} instead of the direct summation.
The adaptive mesh refinement (AMR) method which places hierarchical
meshes where required, is also one of those methods.  Some groups
place rectangular structured mesh \citep{vil89, annnc94, suis95,
gelcw97, norb99} while other groups place non-rectangular structured
mesh \citep{krakk97, knegb01, yahy01, tey02}.  A review of
cosmological $N$-body codes is found in \citet{ber98}.

For the tree method, there are various vectorization schemes
\citep{bar90, her90, mak90} and parallelization schemes for distributed
memory machines \citep{sal90, wars93, dub96, yahmy99, spryw01}.
Moreover, there is a parallel tree code \citep{mak04} accelerated by
the special purpose computer, GRAPE \citep{sugetal90, kawfmt00, makfkn03}.
On the other hand, \citet{fryetal00} discussed the parallelization
scheme of their AMR hydrodynamics code. However, parallelization of
the $N$-body part is not described extensively.  Hence, we describe our
vectorization and parallelization scheme in this paper.  
Some may think that the vectorization is obsolete technique.  However,
the Earth Simulator which had been the fastest computer in the world
until November 2004, is a vector-parallel
type computer, and the vectorized code is easily parallelized for
shared-memory multiprocessors.

In \S 2, we review the data structures and algorithms of our AMR
$N$-body code to define the terminologies used in the following
sections.  In \S 3, we describe the vectorization scheme of the
AMR $N$-body code.  In \S 4, we describe the parallelization scheme of
the AMR $N$-body code, and the timing analysis of the parallelized
code is described in \S 5.  Finally, we summarize this paper in \S 6.

%##################################################
% AMR $N$-body code
%##################################################
\section{AMR $N$-body code} \label{amr:data}
We vectorized and parallelized the shared time step version of the AMR
$N$-body code described in \citet{yahy01}.
We review the AMR $N$-body code briefly in the beginning of this
section.  
Then, since adopted data
structure determines the algorithm suited to the problem, and vice
versa, we review the data structures used in our AMR $N$-body code to
define terminologies used in the following sections, prior to
discussing the vectorization and parallelization schemes.

\subsection{Structure of the code}
The structure of the code is iteration of the following routines:
velocity update stage of the leap-frog integration by a half time
step, position update by a full time step, reconnection of the linked
list of particles, addition and deletion of hierarchical meshes, 
mass assignment from particles to meshes, the Poisson solver, and
velocity update by a half time step.  Force on particles at velocity
update is calculated by interpolation of force on meshes onto
particles.  Position update needs only position and velocity of
particles, because the leap-frog integration is adopted.  After
position update is carried out, some particles go across the cell
boundaries.  Hence, in order to make the linked list consistent,
the linked list of particles are reconnected by picking up particles
which go across the boundaries and inserting them to the linked list
beginning from the finest cells which include them and will not be
deleted at the mesh modification stage.  Mesh generation and deletion
is basically operation using mesh data.  However, when cells are
generated, particles in the parent cells are inherited to the newly
generated cells, and at this stage, particle data are also used.
After this mesh modification, force on mesh is calculated, finally.
First density on meshes are calculated using the position of particles
in the cells.  Then, potential on meshes are calculated from density
on meshes by the Poisson solver.  Force on meshes are calculated by
taking difference of potential at neighboring cells.

We keep the particle linked list consistent as follows:
First, particles are grouped into the stay particles and the leaked
particles.  The stay particles do not move across the boundary of the
cell during the time step, while the leaked particle move across the
boundary.  The leaked particles are removed from the linked list, then
added to a linked list starting from the neighboring cell.  However,
the neighboring cell might be a buffer cell although the starting
cell of the particle linked lists is restricted to a refined cell.
(See \S \ref{vec:mesh} for the definition of a buffer cell.)
In that case, those leaked particles
are passed to the parent cell until they reach to a refined cell or
the base mesh.  On the other hand, the neighboring cell which is to
be connected by the leaked particles may have refined child cells.  As
mentioned above, starting cell of the particle linked list must not
have any child refined cells.  In that case, those leaked particles
are passed to the refined child cell until they reach to the finest
refined cell including them.  At this stage, linked lists are
consistent with the distribution of particles and meshes, but cell
classification and the refinement criteria could be inconsistent.
As the first step of the next stage, cells satisfying the refinement
criteria are flagged, If starting cells of particle linked lists are
not flagged or the buffer child cells are flagged, again, particles in
such cells are passed to the parent cell or the child cell until they
reach to the base mesh or the finest refined cell.  Then, at each
level from the base level to the deepest level, unused cells are
removed and new refined cells are added.  At the same time, particles
connected from the parent cell of the newly refined cells are handed
to the new child refined cells.  Finally, hierarchical meshes are
sorted in the Morton order.  Through these procedures, the linked
lists are consistent with the distribution of particles and meshes and
meshes classification is also consistent with the refinement criteria.

We adopted the Cloud-in-cell (CIC) scheme \citep{hoce88} for force
interpolation and mass assignment.  In the CIC scheme, mass of a
particle is assigned to eight vertices of the cell including the
particle, and force defined at eight vertices of the cell is
interpolated on the particle.  Mass assigned to cells are restricted to
the parent cells by the CIC scheme from the finest mesh level to the
base mesh level.  For the Poisson solver, we use red-black
Gauss-Seidel relaxation.  Further details of our serial AMR $N$-body
code are described in \citet{yahy01}.

As described in the next subsection, particles are connected to the
finest refined cell.  Density at cells which do no have a child cell
octet but include particles are calculated by the CIC scheme.  On the
other hand, density at cells which have a child cell octet is
calculated by the full weighting fine-to-coarse operator from the
child cells and their neighbor cells:
\begin{eqnarray*}
\rho^{L-1}_{2l, 2m, 2n} = &F&(\rho^{L})_{2l, 2m, 2n}\\
=
	&\frac{1}{8}&\rho^{L}_{2l, 2m, 2n} + \\
	&\frac{1}{16}&(\rho^{L}_{2l-1, 2m, 2n} +
			\rho^{L}_{2l+1, 2m, 2n} +
			\rho^{L}_{2l, 2m-1, 2n} +
			\rho^{L}_{2l, 2m+1, 2n} +
			\rho^{L}_{2l, 2m, 2n-1} +
			\rho^{L}_{2l, 2m, 2n+1}) +\\
	&\frac{1}{32}&(
			\rho^{L}_{2l-1, 2m-1, 2n} +
			\rho^{L}_{2l+1, 2m-1, 2n} +
			\rho^{L}_{2l-1, 2m+1, 2n} +
			\rho^{L}_{2l+1, 2m+1, 2n} +\\
		    &&	\rho^{L}_{2l-1, 2m, 2n-1} +
			\rho^{L}_{2l+1, 2m, 2n-1} +
			\rho^{L}_{2l-1, 2m, 2n+1} +
			\rho^{L}_{2l+1, 2m, 2n+1} +\\
		    &&	\rho^{L}_{2l, 2m-1, 2n-1} +
			\rho^{L}_{2l, 2m+1, 2n-1} +
			\rho^{L}_{2l, 2m-1, 2n+1} +
			\rho^{L}_{2l, 2m+1, 2n+1}) +\\
	&\frac{1}{64}&(	
			\rho^{L}_{2l-1, 2m-1, 2n-1} +
			\rho^{L}_{2l+1, 2m-1, 2n-1} +
			\rho^{L}_{2l-1, 2m+1, 2n-1} +
			\rho^{L}_{2l+1, 2m+1, 2n-1} +\\
		    &&	\rho^{L}_{2l-1, 2m-1, 2n+1} +
			\rho^{L}_{2l+1, 2m-1, 2n+1} +
			\rho^{L}_{2l-1, 2m+1, 2n+1} +
			\rho^{L}_{2l+1, 2m+1, 2n+1}).
\end{eqnarray*}
It is possible to prove the following equation:
\begin{eqnarray*}
\rho^{L}_{\mathrm{CIC}}
&=& F(\rho^{L+1}_{\mathrm{CIC}})\\
&=& F(F(\rho^{L+2}_{\mathrm{CIC}}))\\
&\vdots&\\
&=& F^{N}(\rho^{L+N}_{\mathrm{CIC}})\\
&\vdots&
\end{eqnarray*}
where $\rho^{L}_{\mathrm{CIC}}$ is the density field derived by the
cloud-in-cell scheme directly.

\subsection{Data structure used in the code}
\begin{figure}
\FigureFile(80mm,50mm){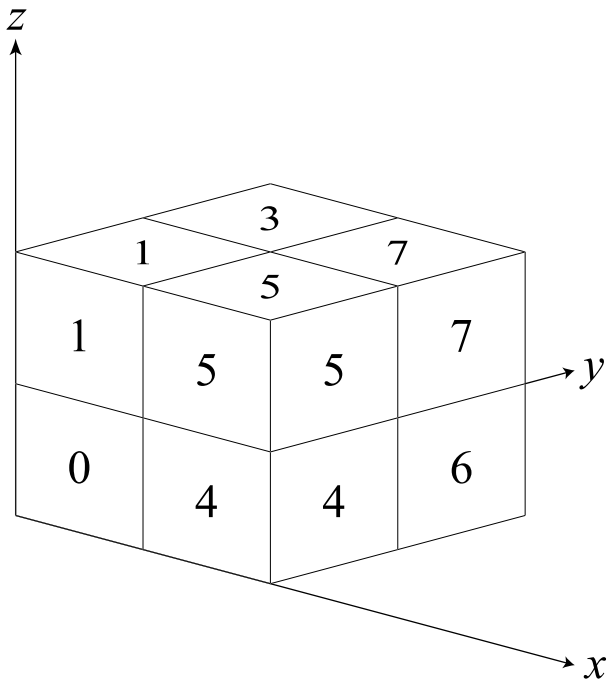}
\caption{
Position of eight cells in a cell octet.  Eight sibling cells are
numbered by the Morton order and their position value is given in
Table \ref{table:cellpos}. }
\label{fig:amrchld}
\end{figure}

\begin{figure}
\FigureFile(80mm, 50mm){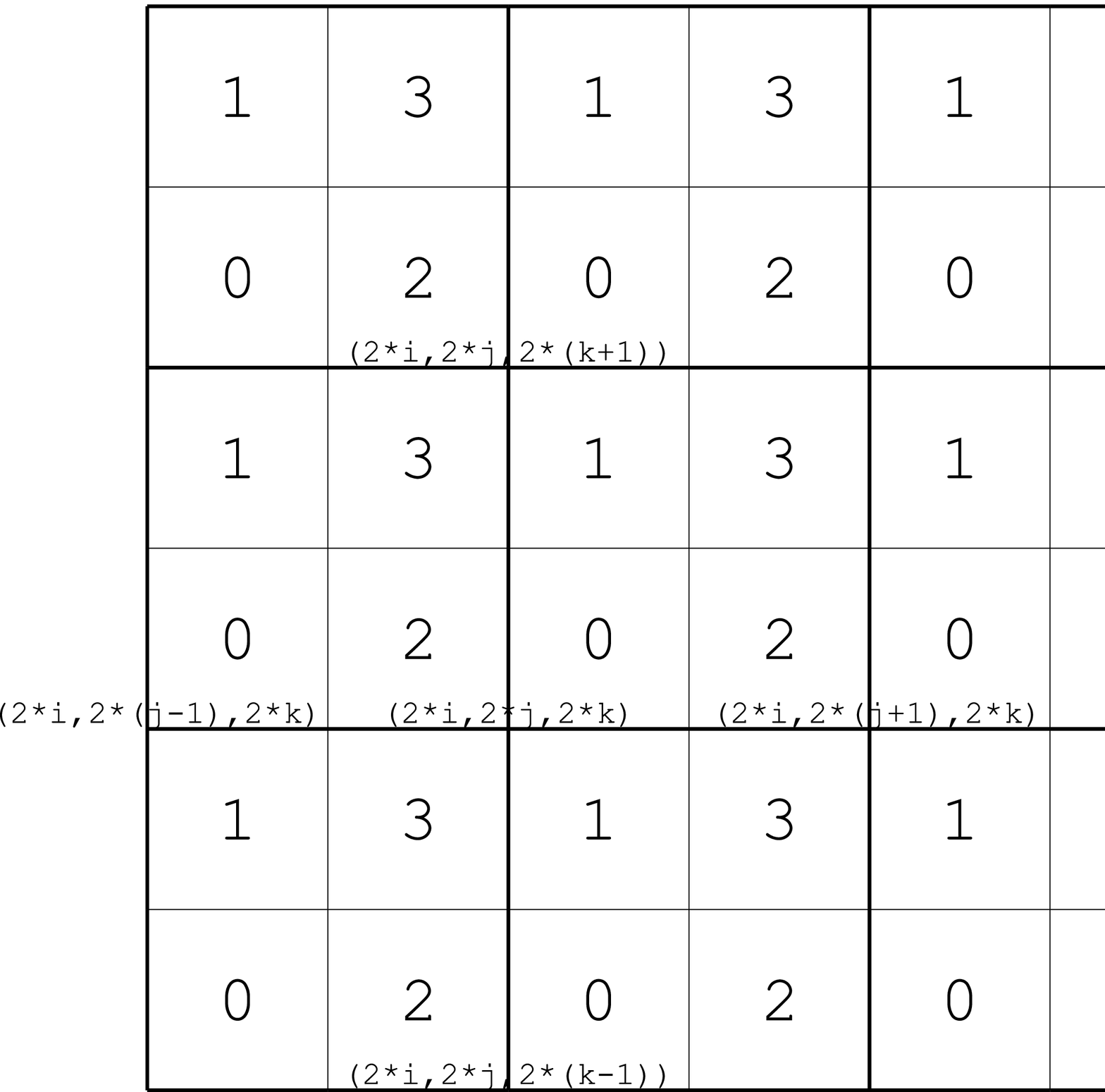}
\caption{
Position of neighboring cells in $y$-direction and $z$-direction of
cells in a cell octet placed at ({\tt 2*i}, {\tt 2*j}, {\tt 2*k}).
Position of cells and cell octets is defined at their lower left
corner.  For example, four of six neighboring cells of the 0-th child
cell in the cell octet placed at the center in this figure are shown
(see also Table \ref{table:nghbpos}). }
\label{fig:amrnghb}
\end{figure}

Next, we describe the data structure used in our code.
First, we divide data into two categories: particle and mesh.  Data
structure of particles is fairly simple.  Particles are just stored in
arrays.  Data structure of meshes is much more complicated. First,
meshes are divided into two groups further: base mesh and hierarchical
mesh.  Data structure of the base mesh is simple.  The base mesh is
stored in arrays.  Accessing neighboring cells at the base mesh is
trivial, if they have a frame, or ghost points.  For example, if the
size of base mesh is {\tt N}$^3$, arrays of {\tt (N+2)}$^3$ are
declared for the base mesh, and inside {\tt N}$^3$ cells are used for
calculation.  Data structure of the hierarchical meshes is more
complicated.  Eight cells refined from the same parent cell are
bundled into a single data structure \citep{khk98}, we call it cell
octet.
Cells and cell octets are also stored in arrays.  Thus, a cell whose
index is \verb+ic+ belongs to a cell octet whose index is
\verb+ic >> N_DIM+, where \verb+N_DIM+ is the number of dimension and
\verb+>>+ represents a right shift operator.  For example, when
\verb+N_DIM=3+, a cell \verb+ic+ belongs to a cell octet \verb+ic/8+.
Since cell octets have indices to the six neighboring cell octets, it
is easy to access the neighboring cell octets from a cell octet.  
However, it is not so simple to access six neighboring cells from a
cell.
First, we
numbered siblings in a cell octet in the Morton order (see \S
\ref{para:datadcmp}).
For example, consider a cell octet placed at ($x$, $y$, $z$) =
({\verb+2*i, 2*j, 2*k+}), where \verb+i+, \verb+j+, and \verb+k+ are
arbitrary integers.  At this time, {\tt l}-th child cell of the cell
octet is placed at ({\verb+(2*i | ((l & 4)>>2),+}
{\verb+2*j | ((l & 2)>>1),+} {\verb+2*k | (l & 1)+}), as given in Table
\ref{table:cellpos} (see also Fig. \ref{fig:amrchld}).
Here, \verb+&+ and \verb+|+ represent a bitwise AND operator and a
bitwise OR operator, respectively.
Then, six neighboring cells of the 0th child are given in Table 
\ref{table:nghbpos}.
It is trivial to extend the above relation to the neighbors of other
seven child cells (see also Fig. \ref{fig:amrnghb}).

\begin{table}
\begin{center}
\caption{Position of {\tt l}-th child cell\label{table:cellpos}}
\begin{tabular}{cccc}
\hline
\hline
{\tt l} & \multicolumn{3}{c}{Position of cell}\\
\hline
{\tt 0} & ( {\verb|2*i  |}, & {\verb|2*j  |}, & {\verb|2*k  |} )\\
{\tt 1} & ( {\verb|2*i  |}, & {\verb|2*j  |}, & {\verb|2*k+1|} )\\
{\tt 2} & ( {\verb|2*i  |}, & {\verb|2*j+1|}, & {\verb|2*k  |} )\\
{\tt 3} & ( {\verb|2*i  |}, & {\verb|2*j+1|}, & {\verb|2*k+1|} )\\
{\tt 4} & ( {\verb|2*i+1|}, & {\verb|2*j  |}, & {\verb|2*k  |} )\\
{\tt 5} & ( {\verb|2*i+1|}, & {\verb|2*j  |}, & {\verb|2*k+1|} )\\
{\tt 6} & ( {\verb|2*i+1|}, & {\verb|2*j+1|}, & {\verb|2*k  |} )\\
{\tt 7} & ( {\verb|2*i+1|}, & {\verb|2*j+1|}, & {\verb|2*k+1|} )\\
\hline

\end{tabular}
\end{center}
\end{table}

\clearpage
\begin{table}
\begin{center}
\caption{Neighbor cells of 0th child cell\label{table:nghbpos}}
\begin{tabular}{ccccccc}
\hline
\hline
\multicolumn{3}{c}{Neighbor's position} &
\multicolumn{3}{c}{Position of cell octet} &
index of cell \\
\hline
( {\verb|2*i-1|}, & {\verb|2*j  |}, & {\verb|2*k  |} ) &
( {\verb|2*i-2|}, & {\verb|2*j  |}, & {\verb|2*k  |} ) & {\tt 4} \\
( {\verb|2*i+1|}, & {\verb|2*j  |}, & {\verb|2*k  |} ) &
( {\verb|2*i  |}, & {\verb|2*j  |}, & {\verb|2*k  |} ) & {\tt 4} \\
( {\verb|2*i  |}, & {\verb|2*j-1|}, & {\verb|2*k  |} ) &
( {\verb|2*i  |}, & {\verb|2*j-2|}, & {\verb|2*k  |} ) & {\tt 2} \\
( {\verb|2*i  |}, & {\verb|2*j+1|}, & {\verb|2*k  |} ) &
( {\verb|2*i  |}, & {\verb|2*j  |}, & {\verb|2*k  |} ) & {\tt 2} \\
( {\verb|2*i  |}, & {\verb|2*j  |}, & {\verb|2*k-1|} ) &
( {\verb|2*i  |}, & {\verb|2*j  |}, & {\verb|2*k-2|} ) & {\tt 1} \\
( {\verb|2*i  |}, & {\verb|2*j  |}, & {\verb|2*k+1|} ) &
( {\verb|2*i  |}, & {\verb|2*j  |}, & {\verb|2*k  |} ) & {\tt 1} \\
\hline

\end{tabular}
\end{center}
\end{table}


There is yet another data structure: the linked list of particles.
Particles in the same cell are connected by a linked list.
If a refined cell whose parent cell satisfies the refinement
criteria does not have any child refined cells, such refined cell
has the index of the head particle of the linked list consisting of
particles in the cell.

%##################################################
% Vectorization
%##################################################
\section{Vectorization}
With pipeline processing, vector processors calculate loops in codes
much faster than usual scalar processors on personal computers and work
stations.  Moreover, there are many vector processors in Japan,
including the supercomputer system of the National Astronomical
Observatory of Japan which is exclusively served to the astronomical
community, and the Earth Simulator.  
Thus, we decided to vectorize our AMR $N$-body code.
In order to bring out the performance of those vector processors,
we must eliminate as many unvectorizable loops in the code as possible.
Most often the loops that are unvectorizable are recursive loops, in
which store operation to a certain address precedes load operation
from the same address during the loop.  We describe the vectorization
schemes of our AMR code in this section.  Hereafter, C-notation is
used for code description unless otherwise mentioned, and some codes
are rewritten for this paper to make clear the argument.

Incidentally, some may think that vectorization is an obsolete
technique.  However the scheme described in this section to eliminate
the unvectorized loops is in common with the parallelization scheme
for shared-memory multiprocessors.

\subsection{Loop types}
Before discussing the vectorization scheme, we classify loops in our
code into three categories from the data structure point of view. The
first category consists of those loops which treat particle data
only. Loops in the integration of particle trajectory are classified
into this group. We call loops categorized into this group particle
loops. These loops are vectorized easily, and usually compilers
vectorize them automatically. The second category includes those loops
which treat mesh data only. We call those loops mesh loops. Loops in
the multigrid Poisson solver are grouped here. Loops treating only the
base mesh are vectorized automatically. Loops treating the
hierarchical meshes need some tricks to be vectorized. However schemes
for those loops are not as elaborate as schemes for loops categorized
into the third group, that treat particles and mesh simultaneously in
the same loop. Mass assignment and force interpolation belong to this
category. We call this group particle-mesh loops. In the rest of this
section, we discuss the vectorization schemes for each type of loops. 

\subsection {Particle loops}
There is only one function which has a particle loop. That is the
position step of the leap-frog time integration of particles. 
Usually, this part is vectorized automatically by a compiler.

\subsection {Mesh loops}\label{vec:mesh}
Vectorization of loops for the base mesh is trivial. 
The mesh loops for the hierarchical meshes is more
complicated than that for the base mesh.  This complexity is partly
due to the data structure mentioned in \S \ref{amr:data}.  
Another source of complexity is the difference of the boundary
condition. The boundary of the base mesh is periodic, while the
boundary of hierarchical meshes is arbitrarily shaped and
non-periodic.  There are two
types of hierarchical meshes.  The first type is refined cells whose
parent cells satisfy the refinement criteria, and the second type is
buffer cells which are placed to wrap the refined cells to provide
boundary conditions for refined cells.  Since
refined cells are always surrounded by other refined or buffer cells,
all potential values assigned to the refined cells are updated at each
iteration of the red-black Gauss-Seidel iteration, for example.
On the other hand, potential is updated only in a fraction of buffer
cells.  
Why is the potential updated in some buffer cells and not others?
Because potential is defined at cell-corners.  Each refined cell octet
has potential of only 8 points, while each cell octet has 27
cell-corners.  If eight neighboring cell octets of a refined cell
octet are also refined cell octets, potential at remained 19 points
are updated by those cell octets.  However, if there are neighboring
buffer cell octets, it is not so simple.  For clarity, consider a one
dimensional case and potential is defined at left boundary of cells.
In this case, when the right neighboring cell of a refined cell is
also refined cell, there is no problem.  However, if the right
neighboring cell is a buffer cell, potential of that buffer cell must
be updated during the iteration of the Poisson solver.  On the other
hand, for a buffer cell whose right neighboring cell is a refined
cell, potential at that buffer cell need not be updated.  Thus,
potential at buffer cells whose left neighboring cell is a refined
cell must be updated, while potential at other buffer cells need not.
Since potential at some buffer cells are updated while potential at
other buffer cells are not, we need flags to indicate which buffer cells
must be updated.  These flags are stored as an integer at each cell as
given in Table \ref{table:pbit}.  Using these flags, the code of the
Poisson solver for the hierarchical meshes is written as in Appendix 
\ref{app:code:psh}.

\begin{table}
\begin{center}
\caption{
Meanings of the nine least significant bits of the index which is used
by the Poisson solver for the hierarchical meshes (See Appendix
\ref{app:code:psh}).
\label{table:pbit}}
\begin{tabular}{lll}
\hline
\hline
macro variable  & value  & meaning\\
\hline
{\tt UPDATE\_ALL  } & {\tt 000000001} & Refined cell. (All must be updated.)\\
{\tt UPDATE0      } & {\tt 000000010} & 0th child must be updated.\\
{\tt UPDATE1      } & {\tt 000000100} & 1st child must be updated.\\
{\tt UPDATE2      } & {\tt 000001000} & 2nd child must be updated.\\
{\tt UPDATE3      } & {\tt 000010000} & 3rd child must be updated.\\
{\tt UPDATE4      } & {\tt 000100000} & 4th child must be updated.\\
{\tt UPDATE5      } & {\tt 001000000} & 5th child must be updated.\\
{\tt UPDATE6      } & {\tt 010000000} & 6th child must be updated.\\
{\tt UPDATE7      } & {\tt 100000000} & 7th child must be updated.\\
{\tt UPDATE\_RED  } & {\tt 011010010} & Updated in the red iteration.\\
{\tt UPDATE\_BLACK} & {\tt 100101100} & Updated in the black iteration.\\
\hline
\end{tabular}
\end{center}
\end{table}

\subsection {Particle-mesh loops}

\begin{figure}
\FigureFile(80mm,50mm){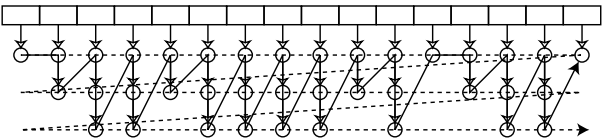}
\caption{
Schematic illustration of the depth-first order sweep and the
breadth-first order sweep. Particles ({\it circles}) in the same cell
({\it rectangle}) are connected by a linked list 
({\it downward arrows}). {\it Vertical solid zigzag} arrow represents
the access pattern of the depth-first order sweep which is
unvectorizable. {\it Horizontal dashed zigzag arrow} represents the
access pattern of the breadth-first order sweep which is vectorizable.}
\label{fig:vecswp}
\end{figure}

It is particle-mesh loops that is most difficult to vectorize among 
three types of loops described in the beginning of this section. In
the scalar
code, particles in a same cell are accessed by rolling up the linked list
(Fig. \ref{fig:vecswp}). The code containing such loops are given in
Appendix \ref{app:code:pms}.
These loops are not vectorizable, because neither the length of the
loops nor which particles are to be accessed are not known
beforehand. This access pattern of particles is called depth-first
order and blocks vectorization frequently.  For example, tree code had
also confronted with this difficulty.  \citet{her90} and \citet{mak90}
overcame this problem by exchanging the depth-first order loops by the
breadth-first order ones.  We followed this strategy.  The code of the
exchanged loops are shown in Appendix \ref{app:code:pmv}.  Most of
particle-mesh loops are vectorized by this order exchange. The
vectorization schemes of three main functions which contain
particle-mesh loops are discussed below.

\subsubsection {Force interpolation}
If each particle has an index of the cell including it, force
interpolation 
is vectorizable even without the order exchange, because only particle
velocity is updated and it is not overlapped in the memory during a
sweep.  However, if each particle does not have an index of the parent
to save the memory usage, the loop order of force interpolation must
be exchanged to be vectorized.

\subsubsection {Mass assignment}\label{vecpar:vec:mass}
In plasma physics, the particle-in-cell (PIC) method is widely used
and the PIC method includes charge assignment which is almost
identical to mass assignment in the PM method.  There is a
vectorization scheme for charge assignment named the Abe-Nishihara
method \citep{nisetal00}.  In the Abe-Nishihara method, particles
assign their charge assuming that particles reside in different
cells, and at the same time, for each cell, the number of particles
which assign their charge to the cell is counted.  If the number of
particles in cells is zero or one, the loop is not recursive.
However, if the number of particles in a cell exceeds one, recursive
memory access has occurred. In that case, it is necessary to correct
the result and the charge is assigned by a scalar loop.  If the loop
length of charge assignment is well short compared with the number of
particles, this correction occurs scarcely.

\begin{figure}
\FigureFile(80mm,50mm){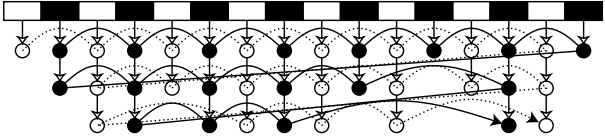}
\caption{
Schematic illustration of the loop splitting for vectorization of the
mass assignment. Adopted the cloud-in-cell mass assignment scheme,
the loop is recursive, if black and white particles assign their mass
simultaneously during breadth-first ordered loops. This is because the
density at a node could be updated more than once during a sweep. This
recursion is avoidable if black particles assign their mass
({\it solid wavy arrow}) after white particles have done
({\it dashed wavy arrow}).}
\label{fig:vecrec}
\end{figure}

However, we developed another vectorization scheme for mass assignment
which does not have neither the scalar parts, nor the correction part,
using the data structure of particles and meshes.
First, if the nearest grid point scheme is adopted for mass
assignment, it is vectorized by the order exchange, because mass of
each particle is assigned to only one cell, and the cells do not
overlap during a sweep.  However, the higher order assignment
schemes, such as the cloud-in-cell scheme or the triangular shaped
cloud scheme, cannot be vectorized because those schemes update more
than one cells data per particle during a sweep of the loop. Such data
accesses are prohibited for vectorization.
However, this difficulty is removed by splitting the loop into
$2^{N_{\mathrm{dim}}}$ loops, where $N_{\mathrm{dim}}$ is the number of
dimension if the cloud-in-cell scheme is adopted.  Figure
\ref{fig:vecrec} shows a one dimensional case. White particles and
black particles assign their mass to cells in the breadth-first order,
separately. Such loops are vectorizable, because density at a node is
updated once per sweep at most.

We classify particles into white and black as follows.  First we
classify cells into black and white. This is possible because cells
having the same parent cell are bundled together and those cells' data
are aligned closely. Thus we can classify the first child cell is
white and the second child cell is black.  Then, particles which are
linked from white cells or black cells are classified as white
particles or black particles, respectively.

\subsubsection {Particle sieve}
We introduced vectorization schemes for particle-mesh loops so far
assuming particles in the same cell are linked listed. However, we
must construct the linked list consistently on our own. Particle sieve
is the function to construct the linked list. First, particle sieve is
divided into two parts: stay particle sieve and leaked particle sieve.
Stay particle sieve checks whether particles remain in the same cell
from the previous step, or leaked into the neighboring cells. Those 
particles which remain are added to the linked list of the cell. This
process is vectorizable if the consistent linked list is prepared at
the previous step. Although, the very first construction of the linked
list is unvectorizable, it is called only once in a run. Leaked
particle sieve, which added the leaked particle to the neighboring
cells, is unvectorizable, either. However, time steps of simulation must
be controlled so that the ratio of the number of leaked particles kept
small, and the CPU time for leaked particle sieve is negligible.

\subsubsection {Effect of vectorization}
How fast is the vectorized code compared with the the scalar code,
especially the particle-mesh loop parts?  For example, as an extreme
case, consider that nothing is done in a particle-mesh loop.  In this
case, because of the overhead due to vectorization, the vectorized
code is slower than the scalar code.  On the other hand, when a lot of
work is done in a particle-mesh loop, the vectorized code is
advantageous.  In order to quantify the advantage of the vectorized
code, we measured the CPU time consumed by the vectorized mass
assignment, particle sieve, and the scalar mass assignment in the case
that there is only a base mesh whose size is $256^3$ and $256^3$
particles are scattered randomly and homogeneously.  The main parts of
the code are given in Appendix \ref{app:svma}.  The measured CPU times
using the VPP5000 are given in Table \ref{table:vec:CPU}.  Even
including the particle sieve part, the vectorized mass assignment is
36.3 times faster than the scalar mass assignment.  Hence, it is 
beneficial to vectorize the particle-mesh loops using the loop order
exchange.

\begin{table}
\begin{center}
\caption{CPU time of functions including a particle-mesh loop
  \label{table:vec:CPU}}
\begin{tabular}{rc}
\hline
\hline
Part & CPU time [s]\\
\hline
vectorized mass assignment (a) \dotfill &  0.48\\
particle sieve (b)             \dotfill &  0.32\\
(a) + (b)                      \dotfill &  0.80\\
scalar mass assignment         \dotfill & 29.04\\ 
\hline

\end{tabular}
\end{center}
\end{table}

%##################################################
% Parallelization
%##################################################
\section{Parallelization}
The fastest computers in the world adopted distributed memory
architecture and have thousands of processors.  Thus, in order to
carry out large scale simulations, simulation codes must be
parallelized for distributed memory machine.  In this section, we
describe parallelization scheme of our AMR $N$-body code.  The scheme
consists of data decomposition and
communication among processors.  Our parallelized code described in
this section uses Message Passing Interface for a communication
library to transfer data among processors.

\subsection {Data decomposition}\label{para:datadcmp}
Data decomposition is the most important part of parallelization. We
must divide data among processors so that computational load is
distributed equally, memory is used uniformly, and communication among
processors is kept minimum.  Adequate data decomposition realizes these
three conditions.  We describe data decomposition of the three types
of data structure in this subsection.

\subsubsection {Base mesh data}
\begin{figure}
\FigureFile(80mm,50mm){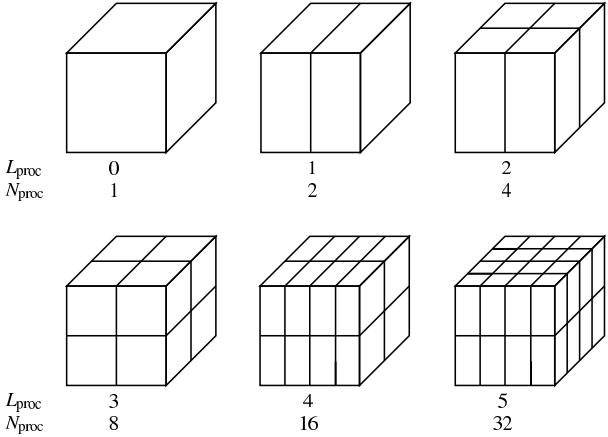}
\caption{
Data decomposition of the base mesh. Here, $N_{\mathrm{proc}}$ is
the number of processors, and $L_{\mathrm{proc}} = \log_2
N_{\mathrm{proc}}$. The shape of the decomposed region
depends on the remainder of $L_{\mathrm{proc}}$ divided by
$N_{\mathrm{dim}}$. In the three dimensional case as the above,
when the remainders are zero, one and two, the decomposed regions are
cubes, square slabs and square pillars, respectively.}
\label{fig:parbase}
\end{figure}

Data decomposition of the base mesh is trivial, because the number
and position of cells are fixed and do not change during a
simulation. Let $L_{\mathrm{proc}}$ be the logarithm of the number
of processors, $N_{\mathrm{proc}}$, to the base 2, that is
\begin{eqnarray*}
L_{\mathrm{proc}} = \log_2 N_{\mathrm{proc}}.
\end{eqnarray*}
Our code assumes that $L_{\mathrm{proc}}$ is integer. Thus, number
of processors is restricted to powers of two.
The shape of the decomposed region depends on the remainder of
$L_{\mathrm{proc}}$ divided by $N_{\mathrm{dim}}$. In the
three dimensional case as shown in Figure \ref{fig:parbase}, when
the remainders are zero, one and two, the decomposed regions are cubes,
square slabs and square pillars, respectively.
Hence, decomposed regions could be elongated. The order of priority
for the elongation is z-direction, y-direction and x-direction,
because we implemented our code in C whose right most indices are
contiguous on the memory. If $L_{\mathrm{proc}}$ is multiple of
$N_{\mathrm{dim}}$, the cubic base mesh whose level is $L_{\mathrm{B}}$ is
decomposed into $N_{\mathrm{proc}}$ cubic base meshes which have
$2^{N_{\mathrm{dim}} \times (L_{\mathrm{B}} - L_{\mathrm{proc}})}$ cells.

\subsubsection {Hierarchical mesh data}
\begin{figure}
\FigureFile(80mm,50mm){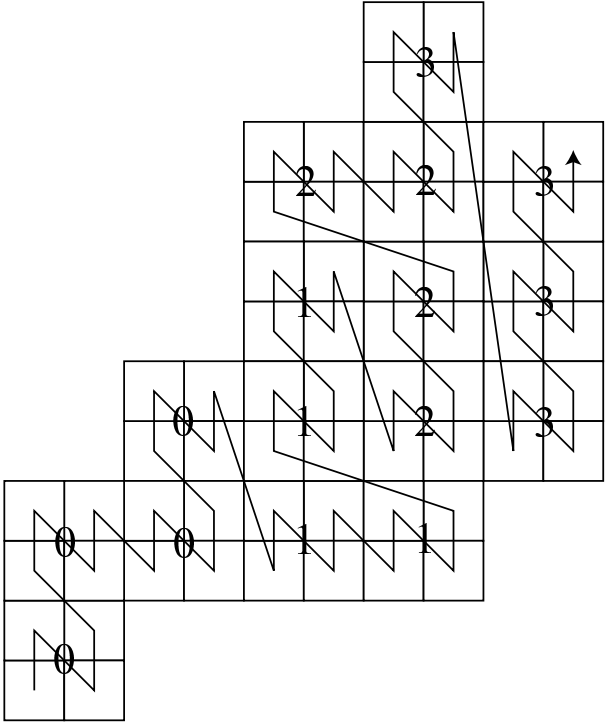}
\caption{
Hierarchical mesh data are sorted in the Morton order 
({\it connected N-shaped arrow}) at each level. Hierarchical meshes
are distributed by dividing the sorted data by the number of
processor, in the above case, four. Numbers in the figure indicate the
ID of the host processor to which the cell octets allocated.}
\label{fig:parmrtn}
\end{figure}

Among three types of data, the hierarchical meshes uses memory most in
our code.  and the most time consuming part of our serial
code is the AMR Poisson solver.  For the $N$-body code, other functions
including particle-mesh loops also use a significant fraction of the
CPU time. However, the
ratio of the CPU time for the mesh loops to the total increases if the
hydrodynamics blocks are incorporated.  Thus, data decomposition of the
hierarchical meshes is the most important part of the
parallelization. Since the red-black
Gauss-Seidel iteration, which is the kernel of the AMR Poisson solver,
is iterated for the hierarchical meshes of the same level, we
should divide hierarchical mesh data equally in each level.
However, we have already sorted the hierarchical meshes in each level by
the Morton order \citep{yahy01}.  Hence, we simply divide
the arrays of hierarchical meshes by $N_{\mathrm{proc}}$ at each
level (Fig. \ref{fig:parmrtn}). Warren \& Salmon (1993) parallelize
the Barnes-Hut tree code using the Morton order. Their code, named hashed
oct tree method, sorts particles in the Morton order and distributes
them to processors. However, we apply the Morton order not to
particles but to the hierarchical meshes.  \citet{fryetal00} also uses
the Morton order for data decomposition.  Difference between our code
and their code is that we sort and divide hierarchical mesh data level
by level, while \citet{fryetal00} sort and divide all hierarchical
meshes at once.  

Our decomposition scheme causes that parent and child cells can be on
different processors.  Hence, communication occurs when accessing child
data from parent cell, or vice versa.  Details of such communication
will be discussed in \S \ref{para:interlevel}.

\subsubsection {Particle data} \label{para:ptcl}
Particles are attached to the finest refined cell including them.
Particles belonging to the same cell are connected by a linked list
and the cell including those particles keeps an index of the first
particle of the linked list.  Hence it is natural that the particles
are assigned to the processor which also covers cells including them.
This data decomposition does not guarantee the balanced
decomposition.  However, if the numbers of cells which contain one
particle, two particles, three particles, etc., are the same among
processors, the numbers of particles in processors are balanced.
We will return to the issue of particle data decomposition in \S
\ref{timana}.

\subsubsection {Example of data decomposition}
\begin{figure}
\FigureFile(80mm,50mm){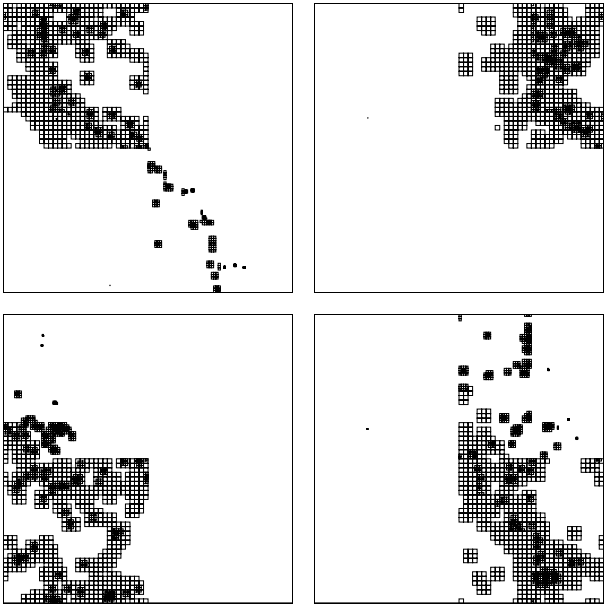}
\caption{
Distribution of the hierarchical meshes among processors. {\it Squares}
indicate cell octets and each panel shows distribution of the cell
octets allocated to the corresponding processor.  For clarity, cells
are placed after projection of particles onto a plane.  Note that some
cell octets and their parent cell octets are assigned to different
processors.}
\label{fig:parexm}
\end{figure}

\begin{figure}
\FigureFile(80mm,50mm){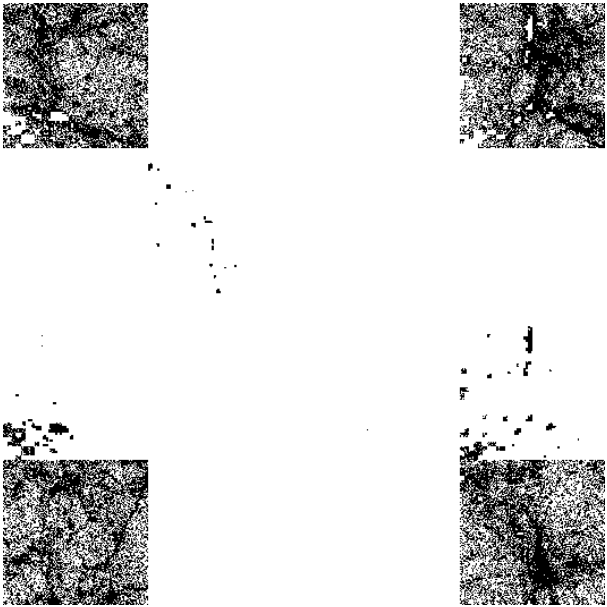}
\caption{Distribution of particles among processors.  Each panel shows 
distribution of the particles allocated to the corresponding
processor.  Distribution of particles is similar to that of the
hierarchical meshes shown in Fig. \ref{fig:parexm}.}
\label{fig:parexp}
\end{figure}

We performed a $\Lambda$-dominated cold dark matter ($\Lambda$CDM)
simulation with $64^3$ particles in the 70 $h^{-1}$ Mpc cubic box
using four processors. The base level and the dynamic range level, or
the deepest level, of the simulation are 6 and 12, respectively.  The
initial condition is prepared by {\tt grafic2} code provided by
\citet{ber01}, but we modified the code to adopt the fitting function
of the transfer function given in \citet{sug95} which includes the
baryon's effect. Figure \ref{fig:parexm} shows the distribution of the
hierarchical meshes at $z=0$ which are assigned to four processors.
Each panel corresponds to a processor.  For clarity, particles are
projected onto a plane, then hierarchical meshes are generated.  Note
that some cell octets and their parent cell octets are assigned to
different processors.  Distribution of particles are shown in Figure
\ref{fig:parexp}.  Distribution of the particles allocated to a
processor is similar to that of the hierarchical meshes allocated to
the same processor.

\subsection {Inter-level communication}\label{para:interlevel}
\begin{figure}
\FigureFile(80mm,50mm){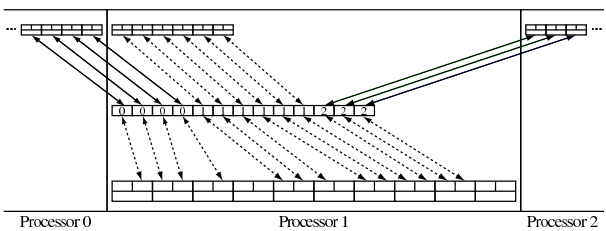}
\caption{
Schematic illustration of the inter-level communication.
Cells ({\it lower rectangles}) collect data from their child cell
octets ({\it upper rectangles}) through the communication buffer
({\it rectangles in the middle}). Solid arrows and dotted arrows
represent inter-processor communication and local load and store,
respectively. Cell octets collect data from their parent cells the
other way around.}
\label{fig:parintlev}
\end{figure}

Since we distribute hierarchical meshes by sorting them at each level,
some cell octets and their parent cells are attached to the different
processors. Therefore, processors must communicate with each other
when data of the child cell octets or parent cells are needed, for
example, when the coarse-to-fine or the fine-to-coarse operators are
called from the AMR Poisson solver. We call such communication among
processors inter-level communication. Figure \ref{fig:parintlev} shows
how inter-level communication is implemented in our code. When cells
({\it lower rectangles}) need data of child cell octets ({\it upper
rectangles}), child cell octets' data are transfered to the
communication buffer ({\it rectangles in the middle}) at the processor
to which parent cells are attached. Then parent cells read data from
the communication buffer. Parent cells' data are transfered to their
child cell octets the other way around. Since cells are sorted in
accordance with the Morton order, it is guaranteed that child cell
octets in the same processor are contiguous on the buffer arrays.

\subsection {Intra-level communication}
Since we simply divide the sorted hierarchical meshes, domain borders
could lie across patches of meshes.  Thus, neighboring cells could be
distributed to different processors after the domain decomposition.
However, the Gauss-Seidel iteration needs potential at six neighboring
cells to update potential at a cell.  Hence, we prepare supplementary
cell octets, or ghost cell octets, to store copies of the neighboring
cell octets distributed to different processors.  After each iteration
of the Gauss-Seidel iteration, data from the original cells are
transfered to the ghost cells by the intra-level communication.

%##################################################
% Timing analysis
%##################################################
\section{Timing analysis} \label{timana}
\begin{figure}
\FigureFile(80mm,50mm){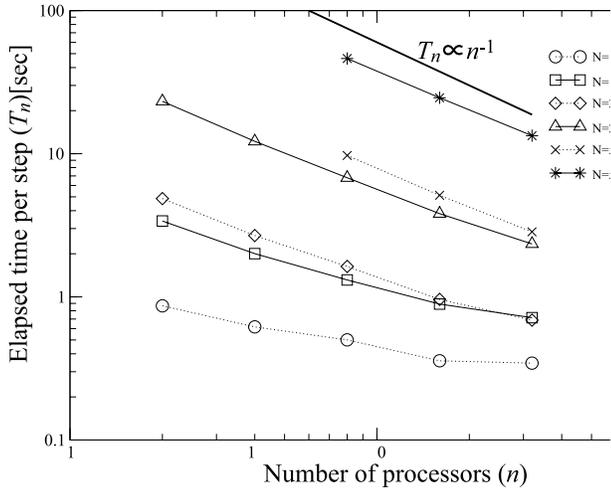}
\caption{
Wall-clock time per step, $T_n$, using $n$ processors.  {\it Circles},
{\it diamonds}, and {\it crosses} represent the wall-clock time measured
during the 38 steps from the beginning of the simulation, using
128$^3$, 256$^3$, and 512$^3$ particles, respectively.  {\it Squares},
{\it triangles}, and {\it asterisks} indicate the wall-clock time measured
during the 40 steps to the end of the simulation, using 128$^3$,
256$^3$, and 512$^3$ particles, respectively.  Thick line shows the
line, $T_n \propto n$.  Slopes of the lines for the case that the
number of particles is 512$^3$, are close to this $T_n \propto n$
line.}
\label{fig:partim}
\end{figure}

We performed three runs of $\Lambda$CDM simulation on a system of
vector parallel processors VPP5000 installed at the Astronomical Data
Analysis Center, National Astronomical Observatory of Japan (ADAC/NAOJ)
to measure the speed of the parallel AMR $N$-body code.  
Each node of VPP5000 has only one CPU and nodes are connected via a
crossbar switch.  Peak performance of each node of VPP5000 is 9.6
GFlops.  Peak data transfer rate is 1.6 GB/s $\times$ 2 (send and
receive).  Since in our parallel code the number of processors is
restricted to be power of two, the number of processors
of the largest queue is 32 in the ADAC/NAOJ system.  Hence we measured
the speed of our code using from 2 processors to 32 processors.
The numbers of particles
of three runs are $128^3$, $256^3$, and $512^3$.  We measured the
wall-clock time per step during 38 steps from the beginning of the
simulation and 40 steps to the end of the simulation ($z=0$) for each
run.  Figure \ref{fig:partim} shows the measured wall-clock times per
step, $T_n$, using $n$ processors.  Thick line in the figure shows the
line, $T_n \propto n^{-1}$.  Slopes of the lines for the case that the
number of particles is 512$^3$ ({\it crosses} and {\it asterisks}),
are close to this $T_n \propto n^{-1}$ line, showing that the speed-up
in that case is close to the ideal speed-up.

On the other hand, parallel efficiency, $p_n$, is widely used as an
indicator of parallel codes, and defined as follows:
\begin{eqnarray*}
p_n = \frac{1}{n}\left(\frac{T_1}{T_n}\right),
\end{eqnarray*}
where $T_1$ and $T_n$ are the wall-clock time using one processor and $n$
processors, respectively.  However, as our case, there are cases that 
$T_1$ is unmeasurable.  In such cases, parallel efficiency is derived
as follows assuming the Amdahl's law:
\begin{eqnarray*}
T_n &=& T_1 (1 - \alpha + \alpha / n)\\
    &=& T_1 (1 - \beta_n \alpha),
\end{eqnarray*}
where $\alpha$ is the parallel fraction, the fraction of time of the 
seral code which is parallelizable, and  $\beta_n=1-1/n$.  Measuring
the wall-clock time using different number of processors, $\alpha$ is
derived as follows:
\begin{eqnarray}
T_n &=& T_1 (1-\beta_n \alpha) \nonumber\\
T_m &=& T_1 (1-\beta_m \alpha) \nonumber\\
\alpha &=&\frac{T_m - T_n}{\beta_n T_m - \beta_m T_n}. \label{eqn:parfrc}
\end{eqnarray}
Then, parallel efficency is:
\begin{eqnarray}
p_n &=& \frac{1}{n}\left(\frac{1}{1-\beta_n \alpha}\right) \nonumber\\
    &=& \frac{1}{n}\left(\frac{\beta_m T_n - \beta_n T_m}{\beta_m T_n
    - \beta_n T_n}\right). \label{eqn:pareff}
\end{eqnarray}

\begin{figure}
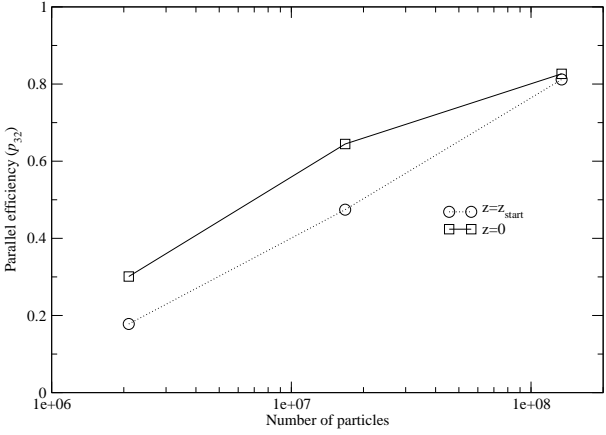

\FigureFile(80mm,50mm){fig11.eps}
\caption{
Parallel efficiency where the number of processor is 32, $p_{32}$,
estimated by Eq. \ref{eqn:pareff} where $m=8$.  {\it Circles} and
{\it squares} indicate the $p_{32}$ at the beginning of the simulation
and the end of the simulation, respectively.}
\label{fig:pareff}
\end{figure}
\begin{figure}
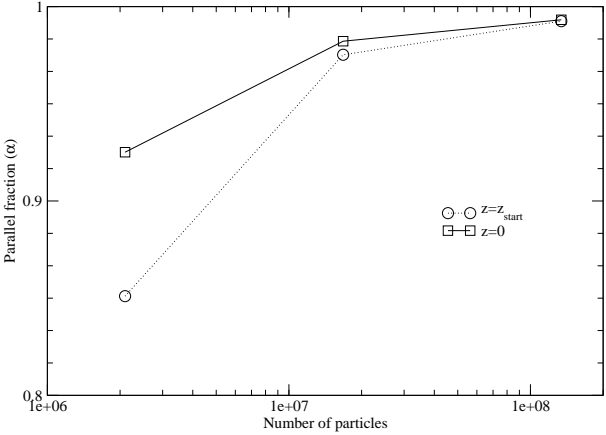

\FigureFile(80mm,50mm){fig12.eps}
\caption{
Parallel fraction estimated by Eq. \ref{eqn:parfrc} where $n=32$ and
$m=8$.  {\it Circles} and {\it squares} indicate the parallel fraction
at the beginning of the simulation and the end of the simulation,
respectively.}
\label{fig:parfrc}
\end{figure}

We estimated the parallel efficiency at $n=32$, $p_{32}$, in the case
that $m=8$.  Figure \ref{fig:pareff} shows $p_{32}$ of each run at
$z=z_{\mathrm{start}}$ ({\it circles}) and $z=0$ ({\it squares}).  Figure
\ref{fig:parfrc} shows the parallel fraction estimated by
Eq. \ref{eqn:parfrc} where $n$=32 and $m$=8.  These two figures
indicate that our parallel code is more efficient as the size of the
simulation increases.

\begin{figure}
\FigureFile(80mm,50mm){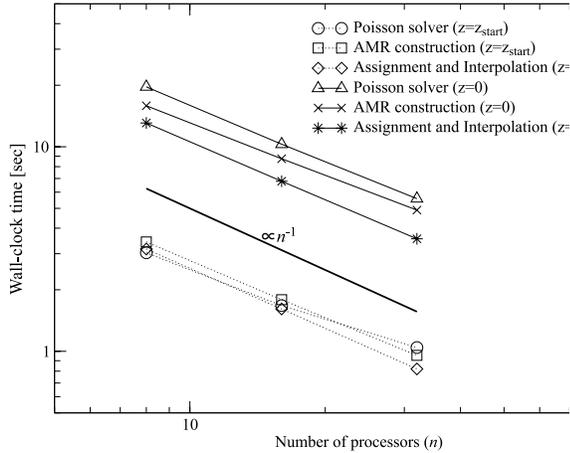}
\caption{
Wall-Clock time of three groups of functions, Poisson solver 
({\it circles} and {\it triangles}), AMR construction including the
reconnection of the linked list of particles ({\it squares} and
{\it crosses}), and mass assignment and force interpolation
({\it diamonds} and {\it asterisks}).  Symbols connected by dotted
lines and solid lines are measured at $z=z_{\mathrm{start}}$ and $z=0$,
respectively.  The number of particles is $512^3$.  {\it Thick line}
indicates the line $\propto n^{-1}$ which shows ideal speed-up.  At the
end of the simulation, wall-clock times of all three groups of functions
shows expected speed-up.  However, at the beginning of the simulation,
speed-up of the Poisson solver slows down.}
\label{fig:parpaa}
\end{figure}

Next, we measured the wall-clock time of three groups of functions:
Poisson solver, AMR construction, and mass assignment and force
interpolation. Figure \ref{fig:parpaa} shows the measured wall-clock
time of each group at $z=z_{\mathrm{start}}$ and $z=0$.  Most groups indicate
the expected speed-up.  However, speed-up of the Poisson solver at
$z=z_{\mathrm{start}}$ slows down.  Since the number of hierarchical meshes at
$z=z_{\mathrm{start}}$ is much smaller than that at $z=0$,  wall-clock time of
the Poisson solver for the base mesh is dominant.  Thus, there might be
room for further optimization of the the Poisson solver for the base
mesh.

\begin{figure}
\FigureFile(80mm,50mm){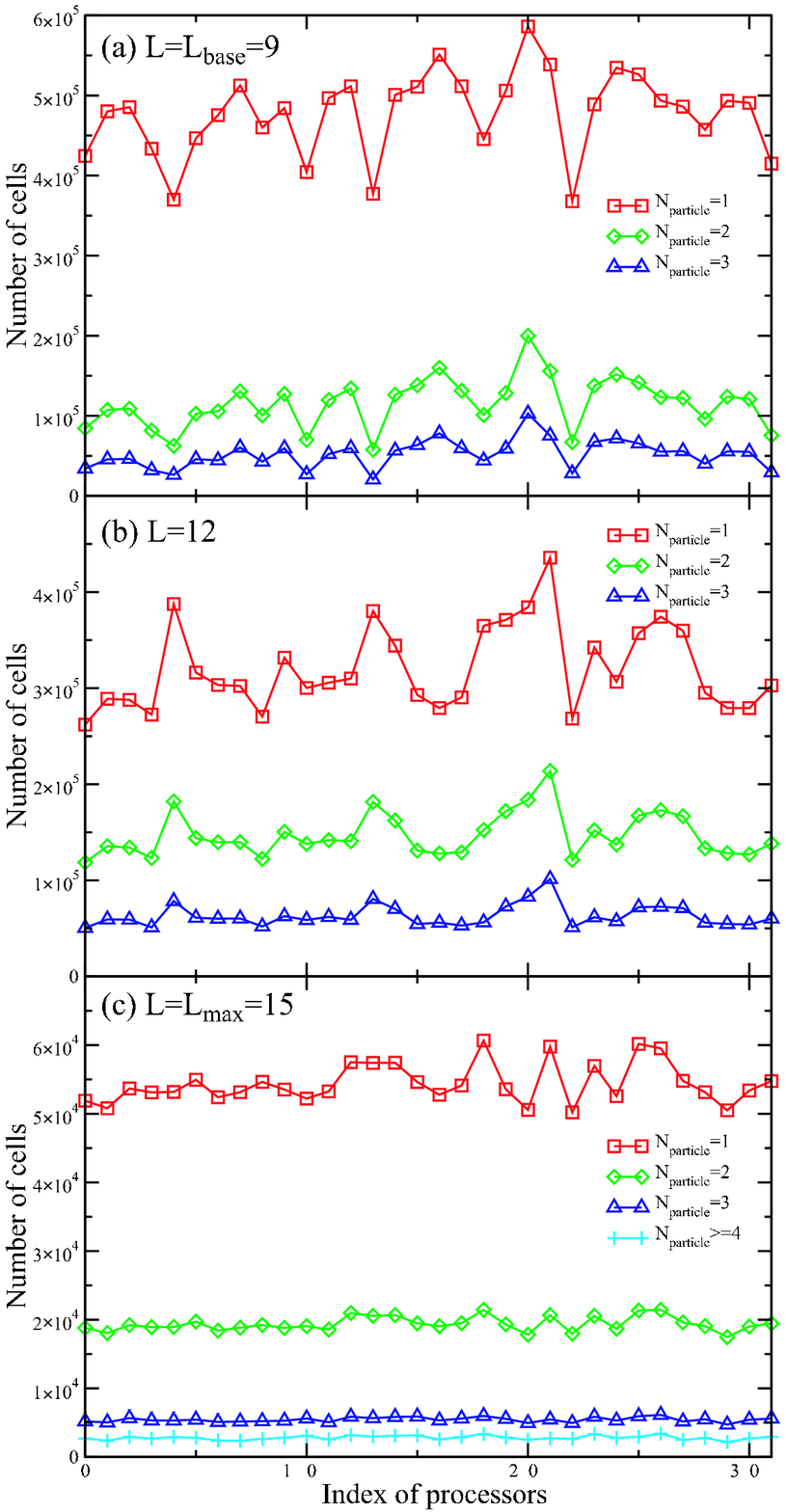}
\caption{
The numbers of cells which contain one particle ({\it squares}), two
particles ({\it diamonds}), three particles ({\it triangles}), and
more than three particles ({\it pluses}) in each processor at the end
of the $\Lambda$CDM simulation.  The number of particles is $512^3$.
Each panel shows the number of cells in (a) the base level ($L=9$),
(b) 12th level, and (c) the finest level ($L=15$).
} 
\label{fig:parncell}
\end{figure}

\begin{figure}
\FigureFile(80mm,50mm){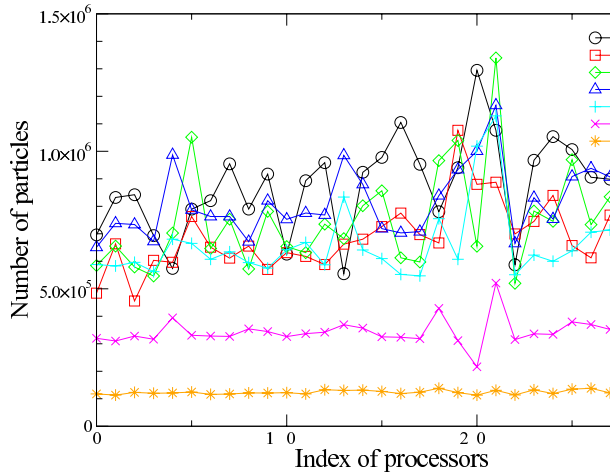}
\caption{
The number of particles assigned to each processor at the end of the
$\Lambda$CDM simulation.  The number of particles in the simulation is
$512^3$.  {\it Circles}, {\it squares}, {\it diamonds},
{\it triangles}, {\it pluses}, {\it crosses}, and {\it asterisks}
indicate the number of particles at levels from 9 (the base level) to
15 (the finest level), respectively.  At the finest level, particles
are distributed in balance, while the dispersion of the number of
particles is large at the coarser levels.
}
\label{fig:parnptcllev}
\end{figure}

\begin{figure}
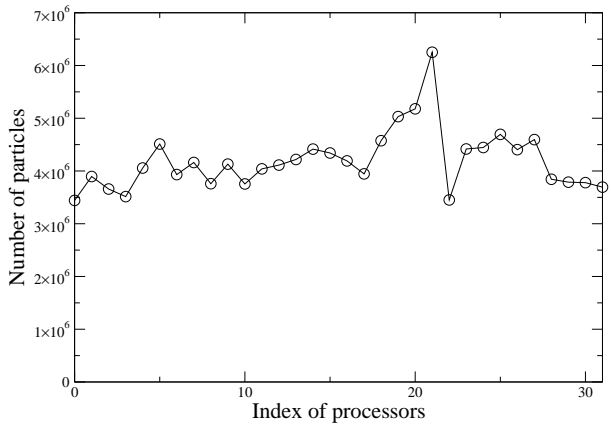

\FigureFile(80mm,50mm){fig16.eps}
\caption{
Total number of particles assigned to each processor at the end of the
$\Lambda$CDM simulation.  The number of particles in the simulation is
$512^3$.  The number of particles in each processor should be
4,194,304, if perfectly balanced.
}
\label{fig:parnptcl}
\end{figure}

As mentioned in \S \ref{para:ptcl}, our data decomposition does not
guarantee the balanced decomposition of particle data.  Figure
\ref{fig:parncell} shows the numbers of cells which include one
particle, two particles, three particles, and more than three
particles at the end of the $\Lambda$CDM simulation using $512^3$
particles.  The numbers of cells including particles is balanced
among processors at the finest level, and the dispersion is large at
the coarser levels.  Figure \ref{fig:parnptcllev} shows the number of
particles in each level among processors.  Again, particles are
equally distributed at the finest level, although cells at the finest
level can contain arbitrarily large number of particles.  Figure
\ref{fig:parnptcl} shows the total number of particles assigned to
each processor.  The processor which is assigned the largest number of
particles has particles about 50\% more than the average number.
However, mass assignment and force interpolation part of our code,
which includes the particle-mesh loops, scales almost ideally as the
number of processors is increased as shown in Figure
\ref{fig:parpaa}.  This is because more mesh loops, which assign cells'
mass to the parent cells, are included into mass assignment and force
interpolation part than particle-mesh loops, which assign particle
mass to the cells including it and interpolate force at cells toward
particles.

%##################################################
% Summary
%##################################################
\section{Summary}
We have vectorized and parallelized the shared time step version of
our AMR $N$-body code.   First, in order to vectorize our code, we
grouped loops into three types, mesh loops, particle loops, and
particle-mesh loops.  It is easy to vectorize mesh loops and particle
loops.  However, vectorization of particle-mesh loops is not so
simple.  Particle-mesh loops contain a hash data structure: an array
of hierarchical meshes and linked lists of particles beginning from
the array.  The simplest way to treat a particle-mesh loop containing
such a hash data structure is the depth-first order loop, but such
loop is unvectorizable.  We vectorized particle-mesh loops by changing
the loop order to the breadth-first order.

We also parallelized our code for distributed memory machines.
First, we decomposed data among processors.  Data decomposition of the
base mesh is trivial.  The hierarchical meshes are sorted by the Morton
order level by level, then split and distributed among processors.
Data of particles are allocated to processors to which the data of
cells including them cells are assigned.  Since the hierarchical
meshes are sorted
level by level, some cell octets and their parent cells could be
assigned to different processors.  Thus we also implemented the
inter-level communication to collect the data from child cell octet
allocated to different processors and vice versa.

Finally, we measured the parallel efficiency of our parallel AMR
$N$-body code for three runs of $\Lambda$CDM simulation.  The parallel
efficiency increases as the number of particle of simulation increases. 
The highest value of the parallel efficiency and the parallel fraction
in the case that the number of processors is 32, are 82.6 \% and 99.3
\%, respectively, and they are measured in the $512^3$ particles
simulation at $z=0$.

With this parallel AMR $N$-body code we carried out large scale
simulations as in \citet{yahny04}, and we are going to investigate the
nature of dark halos of galaxies and clusters of galaxies.

%%%%%%%%%%%%%%%%
%Acknowledgment
%%%%%%%%%%%%%%%%
\bigskip
I would like to thank my thesis advisor Y. Yoshii for lasting
encouragement and valuable suggestions.  I am also grateful to
J. Makino for helpful comments which helped to improve this paper a
lot, and to an anonymous referee for valuable comments. 
Simulations described in this paper were carried out using
Fujitsu-made vector-parallel processors VPP5000 installed at the
Astronomical Data Analysis Center, National Astronomical Observatory
of Japan (ADAC/NAOJ), under the ADAC/NAOJ large-scale simulation
projects (group-ID: yhy35b, rhy15b).  I also acknowledge the support
from the Research Fellowships of the Japan Society for the Promotion
of Science for Young Scientists.

\appendix
\section{Code}

\subsection{Poisson solver for the hierarchical mesh}\label{app:code:psh}

As an example of a mesh loop and in order to show how the flags in
Table \ref{table:pbit} are used, we show the Poisson solver for the
hierarchical meshes here.

\begin{verbatim}
{
  int ico, jco, ic, i0, i1, i2, i3, i4, i5;
  int number_of_refined, number_of_red, number_of_black;
  int *refined_celloct, *red_celloct, *black_celloct;
  double rho0;
  [...]
  
  number_of_refined=0;
  for (ico=head_celloct; ico<tail_celloct; ico++)
    if (celloct[ico].flag & UPDATE_ALL)
      refined_celloct[number_of_refined++] = ico;
  number_of_red=0;
  for (ico=head_celloct; ico<tail_celloct; ico++)
    if (celloct[ico].flag & UPDATE_RED)
      red_celloct[number_of_red++] = ico;
  number_of_black=0;
  for (ico=head_celloct; ico<tail_celloct; ico++)
    if (celloct[ico].flag & UPDATE_BLACK)
      black_celloct[number_of_black++] = ico;
      
  /* Red sweep */
  /* Refined cells */
  for (jco=0; jco<number_of_refined; jco++){
    ico= refined_celloct[jco];
    ic = ico * 8;
    i0 = celloct[ico].neighbor_x_minus * 8;
    i1 = celloct[ico].neighbor_x_plus  * 8;
    i2 = celloct[ico].neighbor_y_minus * 8;
    i3 = celloct[ico].neighbor_y_plus  * 8;
    i4 = celloct[ico].neighbor_z_minus * 8;
    i5 = celloct[ico].neighbor_z_plus  * 8;
    
    cell[ic  ].pot = (cell[i0+4].pot + cell[ic+4].pot +
                      cell[i2+2].pot + cell[ic+2].pot +
                      cell[i4+1].pot + cell[ic+1].pot - 
                      cell[ic  ].rho * rho0) / 6.0;
    cell[ic+3].pot = (cell[i0+7].pot + cell[ic+7].pot +
                      cell[ic+1].pot + cell[i3+1].pot +
                      cell[ic+2].pot + cell[i5+2].pot -
                      cell[ic+3].rho * rho0) / 6.0;
    cell[ic+5].pot = (cell[ic+1].pot + cell[i1+1].pot +
                      cell[i2+7].pot + cell[ic+7].pot +
                      cell[ic+4].pot + cell[i5+4].pot -
                      cell[ic+5].rho * rho0) / 6.0;
    cell[ic+6].pot = (cell[ic+2].pot + cell[i1+2].pot +
                      cell[ic+4].pot + cell[i3+4].pot +
                      cell[i4+7].pot + cell[ic+7].pot -
                      cell[ic+6].rho * rho0) / 6.0;
  }
    
  /* Buffer cells */
  for (jco=0; jco<number_of_red; jco++){
	ico = red_celloct[jco];
    ic  = ico * 8;
	[...]
    if (celloct[ico].flag & UPDATE0){
      cell[ic  ].pot = [...];
    }
    if (celloct[ico].flag & UPDATE3){
      cell[ic+3].pot = [...];
    }
    if (celloct[ico].flag & UPDATE5){
      cell[ic+5].pot = [...];
    }
    if (celloct[ico].flag & UPDATE6){
      cell[ic+6].pot = [...];
        
    }
  }

  /* Black sweep */
  [...]
}
\end{verbatim}

\subsection{Scalar particle-mesh loop}\label{app:code:pms}
\begin{verbatim}
{
  int ip, ico;
  [...]

  for (ico=head_celloct; ico<tail_celloct; ico++){
    for (ip=celloct[ico].head; ip != NULL_PARTICLE; ip=particle[ip].next){
      /* Some operations */
    }
  }
}  
\end{verbatim}

\subsection{Vectorized particle-mesh loop}\label{app:code:pmv}
\begin{verbatim}
{
  int mp, np, jp, ico;
  int *list_of_particle;
  [...]

  np = 0;
  for (ico=head_celloct; ico<tail_celloct; ico++){
    if (celloct[ico].head != NULL_PARTICLE)
      list_of_particle[np++] = celloct[ico].head;
  }
  while (np>0){
    for (jp=0; jp<np; jp++){
      /* Some operations */
    }
    for (jp=0; jp<np; jp++)
      list_of_particle[jp] = particle[list_of_particle[jp]].next;
    mp = np;
    np = 0;
    for (jp=0; jp<mp; jp++){
      if (list_of_particle[jp] != NULL_PARTICLE){
        list_of_particle[np++] = list_of_particle[jp];
      }
    }
  }
}  
\end{verbatim}

\section{Code of the scalar mass assignment and the vectorized mass
  assignment} \label{app:svma}
\subsection{Include file}
\begin{verbatim}
#define N_DIM 3
#define BASE_LEV 8
#define BASE_N (1 <<  BASE_LEV)
#define NBASE  (1 << (BASE_LEV * N_DIM))
#define FBASE_N (BASE_N + 2)
#define NFBASE  (FBASE_N * FBASE_N * FBASE_N)
#define BZIND 1
#define BYIND FBASE_N
#define BXIND (FBASE_N * FBASE_N)

#define NPTCL NBASE
#define NULL_PTCL (-1)

#define MALLOC(x, y, z) \
{\
    x=(y *) malloc (sizeof(y) * (z));\
    if (!x)\
        fprintf(stderr, "malloc failure (%s) in %s: line %d\n",\
                #x, __FILE__, __LINE__);\
}
\end{verbatim}

\subsection{Scalar mass assignment code}
\begin{verbatim}
double rhobm[NFBASE];
double ppos[NPTCL*N_DIM];

void scalar_assign (double *cpusa)
{
    int im, ip, ix, iy, iz, *prnt;
    double ct0, dx, dy, dz, lmdx, lmdy, lmdz;
    double *tpos;

    ct0 = scnd();
    MALLOC(prnt,  int, NPTCL);
    MALLOC(tpos, double, NPTCL * N_DIM);
    
    for (ip=0; ip<NPTCL; ip++){
        ix = (int) ppos[ip*N_DIM  ];
        iy = (int) ppos[ip*N_DIM+1];
        iz = (int) ppos[ip*N_DIM+2];
        prnt[ip] = ix * BXIND + iy * BYIND + iz;
    }

#pragma loop novrec
    for (ip=0; ip<NPTCL*N_DIM; ip++)
        tpos[ip] = ppos[ip] - (double)((int) ppos[ip]);
    
    for (im=0; im<NFBASE; im++) rhobm[im] = 0.0;
    for (ip=0; ip<NPTCL; ip++){
        im = prnt[ip];
        dx = tpos[ip*N_DIM  ]; lmdx = 1.0-dx;
        dy = tpos[ip*N_DIM+1]; lmdy = 1.0-dy;
        dz = tpos[ip*N_DIM+2]; lmdz = 1.0-dz;
        rhobm[im                  ] += lmdx * lmdy * lmdz;
        rhobm[im            +BZIND] += lmdx * lmdy *   dz;
        rhobm[im      +BYIND      ] += lmdx *   dy * lmdz;
        rhobm[im      +BYIND+BZIND] += lmdx *   dy *   dz;
        rhobm[im+BXIND            ] +=   dx * lmdy * lmdz;
        rhobm[im+BXIND      +BZIND] +=   dx * lmdy *   dz;
        rhobm[im+BXIND+BYIND      ] +=   dx *   dy * lmdz;
        rhobm[im+BXIND+BYIND+BZIND] +=   dx *   dy *   dz;
    }
    free(prnt);
    free(tpos);
    *cpusa += scnd() - ct0;
}
\end{verbatim}

\subsection{Vectorized mass assignment code}
\begin{verbatim}
int    hpbm[NFBASE];
int    pnxt[NPTCL];

void vector_assign (double *cpuva)
{
    int ip, im0, im, jm, nm, mm, ix, iy, iz, ich;
    int im0l[] = {0,           BZIND,       BYIND,       BYIND+BZIND,
                  BXIND, BXIND+BZIND, BXIND+BYIND, BXIND+BYIND+BZIND};
    int *lbm, *lpt;
    double ct0, dx, dy, dz, lmdx, lmdy, lmdz;
    double *tpos;
#pragma procedure novrec
    ct0 = scnd();
    
    MALLOC(lbm, int, NBASE);
    MALLOC(lpt, int, NBASE);
    MALLOC(tpos, double, NPTCL * N_DIM);

    for (ip=0; ip<NPTCL*N_DIM; ip++)
        tpos[ip] = ppos[ip] - (double)((int) ppos[ip]);
    for (ich=0; ich<(1<<N_DIM); ich++){
        im0 = BXIND + BYIND + BZIND + im0l[ich];
        nm=0;
        for (ix=0; ix<(BASE_N/2); ix++){
        for (iy=0; iy<(BASE_N/2); iy++){ im = im0 + ix * BXIND + iy * BYIND;
        for (iz=0; iz<(BASE_N/2); iz++, im++){
            if (hpbm[im] != NULL_PTCL) {
                lbm[nm] = im;
                lpt[nm] = hpbm[im];
                nm++;
            }
        }}}
        while (nm){
            for (jm=0; jm<nm; jm++){
                im = lbm[jm];
                ip = lpt[jm];
                dx = tpos[ip*N_DIM  ]; lmdx = 1.0-dx;
                dy = tpos[ip*N_DIM+1]; lmdy = 1.0-dy;
                dz = tpos[ip*N_DIM+2]; lmdz = 1.0-dz;
                rhobm[im                  ] += lmdx * lmdy * lmdz;
                rhobm[im            +BZIND] += lmdx * lmdy *   dz;
                rhobm[im      +BYIND      ] += lmdx *   dy * lmdz;
                rhobm[im      +BYIND+BZIND] += lmdx *   dy *   dz;
                rhobm[im+BXIND            ] +=   dx * lmdy * lmdz;
                rhobm[im+BXIND      +BZIND] +=   dx * lmdy *   dz;
                rhobm[im+BXIND+BYIND      ] +=   dx *   dy * lmdz;
                rhobm[im+BXIND+BYIND+BZIND] +=   dx *   dy *   dz;
            }
            for (jm=0; jm<nm; jm++) lpt[jm] = pnxt[lpt[jm]];
            mm=nm; nm=0;
            for (jm=0; jm<mm; jm++)
                if (lpt[jm] != NULL_PTCL){
                    lbm[nm] = lbm[jm];
                    lpt[nm] = lpt[jm];
                    nm++;
                }
        }
    }
    
    free(lbm);
    free(lpt);
    free(tpos);
    *cpuva += scnd() - ct0;
}
\end{verbatim}
\subsection{Particle sieve code}
\begin{verbatim}
void particle_sieve (double *cpus)
{
    int im, jm, nm, mm, ip, ix, iy, iz, ilk, nlk;
    int *lbm, *lpt, *prnt, *tnxt, *llk;
    double ct0;

#pragma procedure novrec
    ct0 = scnd();
    MALLOC(lbm, int, NBASE);
    MALLOC(lpt, int, NBASE);
    MALLOC(prnt,int, NPTCL);
    MALLOC(tnxt,int, NPTCL);
    MALLOC(llk, int, NPTCL);
    

    for (ip=0; ip<NPTCL; ip++){
        ix = (int) ppos[ip*N_DIM  ];
        iy = (int) ppos[ip*N_DIM+1];
        iz = (int) ppos[ip*N_DIM+2];
        prnt[ip] = ix * BXIND + iy * BYIND + iz;
    }
    nm=0;
    for (im=0; im<NFBASE; im++)
        if (hpbm[im] != NULL_PTCL){
            lbm[nm] = im;
            lpt[nm] = hpbm[im];
            nm++;
        }
    for (im=0; im<NFBASE; im++) hpbm[im] = NULL_PTCL;

    nlk=0;
    while (nm){
        for (jm=0; jm<nm; jm++) tnxt[jm] = pnxt[lpt[jm]];
        for (jm=0; jm<nm; jm++){
            im = lbm[jm];
            ip = lpt[jm];
            if (prnt[ip] == im){
                pnxt[ip] = hpbm[im];
                hpbm[im] = ip;
            }
        }
        for (jm=0; jm<nm; jm++)
            if (prnt[lpt[jm]] != lbm[jm]) llk[nlk++] = lpt[jm];
        mm=nm; nm=0;
        for (jm=0; jm<mm; jm++)
            if (tnxt[jm] != NULL_PTCL){
                lbm[nm] = lbm [jm];
                lpt[nm] = tnxt[jm];
                nm++;
            }
    }

#pragma loop scalar
    for (ilk=0; ilk<nlk; ilk++){
        ip = llk[ilk];
        im = prnt[ip];
        pnxt[ip] = hpbm[im];
        hpbm[im] = ip;
    }

    free(lbm);
    free(lpt);
    free(prnt);
    free(tnxt);
    free(llk);
    *cpus += scnd() - ct0;
}
\end{verbatim}

\end{document}